\documentclass[conference]{IEEEtran}
\IEEEoverridecommandlockouts
\usepackage{cite}
\usepackage{amsmath,amssymb,amsfonts}
\usepackage{algorithmic}
\usepackage{graphicx}
\usepackage{textcomp}
\usepackage{xcolor}
\def\BibTeX{{\rm B\kern-.05em{\sc i\kern-.025em b}\kern-.08em
    T\kern-.1667em\lower.7ex\hbox{E}\kern-.125emX}}
\begin{document}

\title{Spatially Adaptive Detection for Satellite-based QKD under Atmospheric Turbulence Channel

}

\author{\IEEEauthorblockN{Yaoxuan Yang}
\IEEEauthorblockA{\textit{IDCOM, School of Engineering} \\
\textit{The University of Edinburgh}\\
Edinburgh, UK \\
s2673056@ed.ac.uk}
\and
\IEEEauthorblockN{Ivi Afxenti}
\IEEEauthorblockA{\textit{IDCOM, School of Engineering} \\
\textit{The University of Edinburgh}\\
Edinburgh, UK \\
iafxenti@ed.ac.uk}
\and
\IEEEauthorblockN{Majid Safari}
\IEEEauthorblockA{\textit{IDCOM, School of Engineering} \\
\textit{The University of Edinburgh}\\
Edinburgh, UK \\
majid.safari@ed.ac.uk}

}

\maketitle

\begin{abstract}
Quantum key distribution (QKD) provides information-theoretic security and satellite-based quantum key distribution (SatQKD) has demonstrated the potential to extend this communication security to intercontinental scales. However, atmospheric turbulence induces significant distortion in the spatial distribution of received optical beams, while background noise remains approximately uniform across the detector plane. As a result, single-element qubit (quantum bit) detection can be frequently dominated by noise due to the random spatial pattern of the imaged wavefront, thereby degrading the system performance. To address this limitation, we propose to exploit the spatial degrees of freedom of single-photon detector arrays to reject the excessive noise while adapting to channel variations induced by turbulence. We develop a threshold-based selection method that only activates detector elements that have higher probability of registering qubits. We evaluate the performance of the proposed noise-rejection QKD schemes using Monte Carlo simulations considering the impact of diffraction and atmospheric turbulence on the transmitted optical beam in the presence of background and dark noise. The results show that, compared to conventional schemes, the proposed noise-rejection strategy effectively reduces the quantum bit error rate (QBER) and improves the secret key rate (SKR) performance, while the performance gains depend on the turbulence condition. These findings demonstrate the potential of adaptive array receiver design to enhance the robustness of the SatQKD system under realistic atmospheric conditions.

\end{abstract}


\section{Introduction}
Conventional cryptosystems are based on computational assumptions, and this fragility has proven to be a threat for both the present and the future. In contrast, quantum key distribution (QKD), based on fundamental principles of quantum physics, can provide information-theoretic security, ensuring long-term security even in the presence of future advances in computational capability \cite{b1}. Currently, the most mature physical implementation of QKD is based on optical fiber transmission. However, in fiber transmission, signal attenuation increases exponentially with distance, becoming a major bottleneck that limits the achievable transmission range \cite{b2,b3}. To overcome this limitation, satellite-based QKD (SatQKD) over free-space links provides a viable solution. In satellite links, the majority of propagation occurs in vacuum, significantly reducing channel loss and enabling long distance transmission \cite{b4}. This has been experimentally validated for intercontinental QKD transmission \cite{b5}. Nevertheless, SatQKD still remains vulnerable to channel impairment caused by factors such as atmospheric turbulence \cite{b6} and background noise \cite{b5}, which degrades the performance of the system in terms of the quantum bit error rate (QBER) and the secret key rate (SKR) \cite{b7}. In particular, atmospheric turbulence causes strong optical wavefront distortion and intensity fluctuation, which pose some challenges for efficient photon detection. Therefore, more research is required on the robust transmission and receiver strategy for free-space QKD systems.

Existing free space QKD systems commonly use single photon detectors (SPD) such as single photon avalanche diode (SPAD) to detect qubits \cite{b5}. %This device integrates the field on the receiver plane and converts the spatial distribution into a one-dimensional counting process. 
Under the combined effect of atmospheric turbulence and diffraction, the optical field at the focal plane exhibits a speckle pattern. To collect a larger fraction of the optical power, a detector with an increased effective area is required. \cite{b8}. 
However, a larger SPD also increases the collected background noise that is uniformly distributed across the detection area, while it cannot distinguish between the detection regions dominated by signal or noise.  %averages these fluctuations, preventing the receiver from distinguishing signal-dominated and noise-dominated regions. Since turbulence induces a highly non-uniform spatial distribution of signal photons while background noise remains approximately uniform, a single-element receiver inevitably collects noise from regions with low signal contribution, thereby degrading system performance.\cite{b11}

To overcome this limitation, we propose spatially resolved detection using SPD array receivers to improve overall signal to noise ratio. Unlike a single SPD, the array includes several SPD microcells that can independently collect parts of the imaged optical field \cite{b10}. Therefore, the spatial degrees of freedom of the array can be exploited using a selective detection method, allowing us to only activate the signal-dominant region and suppress the noise. Existing research on array-based free space QKD receiver in the literature is limited. Moreover, in those limited cases, detector array elements are typically modeled to be statistically identical, which means they share the same distribution statistically, or are uniformly distributed \cite{b4, b12, b26}. This assumption ignores the strong inhomogeneity of the imaged optical field caused by atmospheric turbulence, which leads to different statistical characteristic across different elements. As a result, the spatial diversity introduced by the array detector is not fully utilized and the receiver may operate at a suboptimal performance level. In classical free-space optical communication literature,  there is more research on array receiver designs which also consider nonuniform spatial distribution of the optical field ~\cite{b11}\cite{b13}\cite{b22}. However, these works are mainly analyzed under high photon flux regime, where receiver can accurately characterize the output by a continuous intensity. While QKD operates at the single photon level and the detection output is a binary variable. As a result, receiver design strategies developed for classical optical communication cannot be applied directly to QKD systems \cite{b3}. 

This paper proposes an adaptive receiver framework to exploit the spatial degrees of freedom of a multi-SPD array receivers for noise suppression in SatQKD systems. The main contributions of this work are as follows. First, we characterize the spatial statistical properties of the intensity at the imaging plane of free-space optical receiver across detector array elements under atmospheric turbulence. Second, we propose an adaptive element activation strategy that takes advantage of the array spatial degrees of freedom and the random dynamic of the speckle pattern to improve the receiver performance. Third, we use realistic turbulent free-space optical channel realizations generated based on split-step Fourier method and Monte Carlo simulation to demonstrate that the proposed method can achieve notable improvements in QBER and SKR under a range of turbulent channel conditions.
\section{System Model}
This section presents the system model for the underlying satellite-based QKD link. The system model for optical propagation through the atmospheric link and receiver structure is introduced first, followed by the spatial intensity fluctuations induced by turbulence. These models lay the foundation for the proposed detection scheme.
\subsection{System Description}
We consider a satellite-to-ground downlink QKD scenario in which the quantum signal is transmitted to a ground receiver through a free-space optical (FSO) link. Most of the propagation occurs in vacuum, with only \(8\) km near the ground affected by atmospheric turbulence \cite{b14}. At the receiver, the optical field is collected by a circular aperture with a diameter of \(0.5\,\mathrm{m}\) and is focused onto the focal plane using a thin lens. The intensity distribution is determined by the combined effects of diffraction and turbulence, leading to wavefront distortion \cite{b15}. A two-dimensional detector array equipped with $N$ SPDs is placed at the focal plane of the receiver lens, enabling the focal plane to be resolved into multiple parallel detection channels.

We set \({I}_i\) as the optical intensity received by the \(i\)th detector element. Due to the effect of turbulence, \({I}_i\) shows significant fluctuations, indicating the stochastic nature of the channel. In addition to the signal photon, each detector element is also affected by noise signals such as background noise and dark count noise \cite{b16}, which are assumed to be independent across elements. The joint effect of signal and noise on each detector element will determine the overall output of the array, affecting the performance of the system.

\subsection{Statistical Model of Atmospheric Channel}
The atmospheric channel can be considered quasi-static within each clock period, since the turbulence coherence time is on the order of milliseconds, which is much longer than the clock period of the transmitter which is at the nanosecond level \cite{b15}. The transmitted field is modeled as a Gaussian beam. The majority of the propagation path is vacuum, i.e., the beam is only affected by diffraction. Turbulence is introduced in the lower atmosphere by applying split-step random phase screens along the path \cite{b25}. This split step Fourier method (SSFM) models the atmospheric turbulence by introducing random phase distortion which follows Kolmogorov statistics along the propagation path to stimulate the cumulative effect of refractive index fluctuations. The inner and outer scales of the turbulence are assumed to be \(l_0=5\) mm and \(L_0=20\) m. The random phase screens are placed at every \(50\) m and generated based on the modified von Karman spectrum:\begin{equation}
\Phi(\kappa)=\beta_1 C_n^2\left[1+\beta_2\left(\kappa / \kappa_l\right)-\beta_3\left(\kappa / \kappa_l\right)^{7 / 6}\right] \frac{\exp \left(-\kappa^2 / \kappa_l^2\right)}{\left(\kappa_0^2+\kappa^2\right)^{11 / 6}},
\end{equation}
where $\beta_1=0.033, \beta_2=1.802, \beta_3=0.254, \kappa_l=3.3 / l_0$, $\kappa_0=2 \pi / L_0$, and $C_n^2$ is the refractive index structure constant. The refractive index structure constant is obtained from Hufnagel-Valley (HV) model, given by% Specifically, the HV 5/7 model is adopted. Which is given by: % Requires: \usepackage{amsmath}
\begin{equation}
\begin{aligned}
C_n^2(h) =\; 
& 0.00594 \left( \frac{v}{27} \right)^2 (10^{-5} h)^{10} e^{-h/1000} \\
& + 2.7 \times 10^{-16} e^{-h/1500} \\
& + A e^{-h/100},
\end{aligned}
\label{eq:HV57}
\end{equation}
where \(v\) is the high altitude wind speed, \(A\) is the ground level turbulence strength parameter. By adjusting the parameters, different turbulence conditions from weak to strong are considered. The high altitude wind speed \(v\) is set as \(21\) m/s. The ground level turbulence strength parameter is set to be \(1.7\times10^{-14}\) under weak turbulence, \(1.7\times10^{-13}\) under moderate turbulence, and \(1.7\times10^{-11}\) under strong turbulence. The total distance from satellite to the ground receiver is \(1200\,\mathrm{km}\). In this model, the turbulence effect is assumed to exist only within the lowest \(8\,\mathrm{km}\) of the atmosphere, while the remaining \(1192\,\mathrm{km}\) propagation is in vacuum \cite{b24}. Note that factors such as excessive absorption and scattering due to adverse weather conditions and misalignment are not considered in the study. In addition, the overall path loss and therefore the channel transmittance can be obtained directly from numerical integration of collected wavefront at the receive aperture. 

\subsection{Normalized Intensity Model}
Assuming an ideal single-photon source at the transmitter, a maximum of one signal photon can arrive at the SPD array within one clock period. Under the condition of the signal photon arriving at the detector, the detection process can be considered as determining the spatial location of the signal photon being detected. The probability of a photon being detected in a specific location is determined by the intensity distribution of the optical field in a classical sense. The intensity received \(I_i\) in the \(i\)th element is proportional to the photon arrival rate of the element \cite{b21}. Therefore, during qubit transmission, the probability that a signal photon arriving at the array is detected by the \(i\)th  element \(P_i\) can be expressed as:\begin{equation}
    P_i = \frac{I_i}{\sum_{j=1}^{N} I_j}.
    \label{eq:placeholder_label1}
\end{equation}

\subsection{Noise Model}
The noise affecting each detector element consists mainly of background radiation and intrinsic noise such as dark count~\cite{b3}. In SatQKD, it can be assumed that the background noise within the FOV is spatially uniform between elements~\cite{b22}. In practice, dark count rates vary slightly across elements. However, for analytical tractability and since background noise typically dominates in SatQKD, dark counts are modeled as an independent identical distribution between elements as well \cite{b23}. Under this assumption, the total noise counts within one clock period can be modeled as a Poisson random variable~\cite{b16}. Furthermore, due to the low photon count rate in SatQKD, other SPD imperfections such as the effect of dead time is neglected. The noise level, which is the average number of noise photons received by each element, consists of the background noise and dark count. To evaluate performance under different noise conditions, multiple noise levels are considered. The noise levels are set as \(10^{-8}\), \(5\times10^{-8}\), \(10^{-7}\), \(5\times10^{-7}\) and \(10^{-6}\)  \cite{b5} \cite{b12}.

\section{Noise-Rejection Detection strategy}
Due to the quasi-static nature of the turbulent channel, we can assume that the transmittance \(\eta\) and spatial distribution \(P_i\) remain constant within one coherence time, while across different coherence times, the channel experiences independent realizations, resulting in different values of transmittance and spatial distribution. Therefore, the problem can be formulated as follows: for each realization, the receiver observes the spatial distribution and adjusts its detection strategy accordingly. To this end, only a subset of SPDs is activated for qubit detection at each realization in order to reject the noise of SPDs where the signal is dominated by noise. We denote \({\mathbf S}\subseteq\{1,2,...,N\}\) with cardinality $K$ as the set of activated SPDs. Then the objective is to find the activation set $\mathbf S$ that optimizes the performance of the QKD system.% secret key rate. This is achieved by using the spatial distribution and transmittance to enhance the contribution of signal photon and mitigate the effect of noise. and \({\mathbf P_{\bf S}}=\{P_i: i\in \mathbf S\}\) as the corresponding set of spatial distribution values

\subsection{QKD Performance Eveluation Metrics}
We consider a standard BB84 protocal. The receiver consists of four SPD arrays corresponding to two measurement bases~\cite{b12}. We assume ideal photon preparation and perfect basis alignment, such that a signal photon is detected only by the correct array when the bases match. 

A valid detection event is defined as a single click event, which means that exactly one array is activated within the clock period. Multiple detections within the same array are counted as a single event. The single click event can be caused by either the signal photon clicking the array and the other three arrays failing to detect any noise photon, or, in the case where the array fails to detect a signal photon, but one of the four arrays detects noise. A sifted bit event is defined when the measurement bases chosen are matched and a single click event happens. Using similar approach to the previous study \cite{b12} and considering the activated SPD set of $\mathbf S$ at all four detector arrays, the sifted bit rate (SBR) can be written as% Requires: \usepackage{amsmath}
\begin{equation}
\label{eq:placeholder_label2}
\begin{aligned}
\mathrm{SBR}(\mathbf S) = \frac{1}{2} P_{\mathrm{tx}} \Big[
& \eta \left( e^{-K n_{\mathrm{B}}} \right)^{3} \sum_{i\in \mathbf S} P_{i} \\
& + 4 \left( 1 - \eta \sum_{i\in \mathbf S} P_{i} \right)
\left( e^{-K n_{\mathrm{B}}} \right)^{3}
\Big],
\end{aligned}
\end{equation}
where \(P_{tx}\) is the transmission rate at the transmitter, \(\eta\) is the transmittance of the transmission channel, \(K\) is the number of elements selected by the adaptive receiving strategy, \(n_B\) is the noise phton rate for each SPD, \(P_i\) is the spatial probability of the \(i\)th element. Note that the factor of \(\frac{1}{2}\) corresponds to the probability of selecting the same measurement basis. In addition, the terms $e^{-K n_{\mathrm{B}}}$ and $\eta \sum_{i\in \mathbf S} P_{i}$ represent the probabilities that the activated part of the array does not register a noise photon and that it does not register the signal photon, respectively. 

Under the assumption of perfect system alignment, errors can only be induced by the noise photon. An error event will occur when exactly one array detects a background noise or dark count and that no signal photon is detected. The error bit rate (EBR) can now be written as 
\begin{equation}
\label{eq:placeholder_label3}
\begin{aligned}
\mathrm{EBR}(\mathbf S) = \frac{1}{4} P_{\mathrm{tx}} \Big[4 \left( 1 - \eta \sum_{i\in \mathbf S} P_{i} \right)
\left( e^{-K n_{\mathrm{B}}} \right)^{3}
\Big].
\end{aligned}
\end{equation}
The factor now becomes \(\frac{1}{4}\) because for sifted events caused by noise, only half of them will contribute to noise. The QBER can be written as the ratio of EBR to SKR. % Requires: \usepackage{amsmath}
\begin{equation}
\label{eq:placeholder_label}
\begin{aligned}
\mathrm{QBER}(\mathbf S) =\frac{{\rm EBR}}{{\rm SBR}}.
\end{aligned}
\end{equation} 

Although QBER measures transmission reliability, it cannot fully reflect the efficiency of secret key generation in the QKD system. In practice, the truly decisive metric is the secret key rate (SKR), which quantifies the rate at which the secret key is generated. The SKR depends on both the efficiency of generating sifted bit event and the error rate. According to principles of information theoretic security, the SKR is defined as the difference between the information shared by the legitimate parties and the information potentially accessible to the eavesdropper. In a practical system, the SKR can be written in terms of SBR and QBER \cite{b3} as % Requires: \usepackage{amsmath}
\begin{equation}
    \mathrm{SKR}(\mathbf S) = \mathrm{SBR(\mathbf S)} \left( 1 - 2{\rm H}_{2}(\mathrm{QBER(\mathbf S)}) \right), \label{eq:placeholder_label4}
\end{equation}
here, \({\rm H}_2(.)\) denotes the binary entropy function. It can be observed that both SKR and QBER are directly affected by the subset of selected SPDs, \({\mathbf S}\) and its cardinality \(K\). %Hence the selection strategy will determine the contribution of signal and noise that affect the SKR.  

%\subsection{Limitation of QBER}
To better understand the role of different performance metrics, we evaluated the behavior of QBER as a function of \({\mathbf S}\). Denoting QBER of the best activation set with cardinality $k$ as \({\rm QBER}_k\). To find the best \({\mathbf S}\), we should find the minimum value of $k$, where the inequality \({\rm QBER}_{k+1}/{\rm QBER}_{k}\geq1\) holds. Note that the best activation set with cardinality $k$ should include the SPDs with highest spatial distribution values, $P_i$.  Assuming that the \(P_i\) values are arranged in decreasing order, the inequality can be expressed as % Requires: \usepackage{amsmath}
\begin{equation}
    \frac{e^{-k n_{\mathrm{B}}} - e^{-(k+1)n_{\mathrm{B}}}}{1 - e^{-k n_{\mathrm{B}}}} \sum_{i=1}^{k} P_i \;\ge\; P_{k+1}.
    \label{eq:placeholder_labe5l}
\end{equation}
Assuming \(n_B\) is much smaller than 1, we can take the approximation \(e^{-k n_B}\approx1-kn_B\), then we get \(\frac{1}{k}\sum_{i=1}^{k}P_i\geq P_{k+1}\). Since \(P_i\) values are in decreasing order, this approximated inequality always holds. This implies that optimizing towards minimum QBER would tend to select a very limited number of detector elements, or even only the one with the highest probability. However, such a selection strategy significantly decreases SBR and thereby SKR, although it minimizes QBER. Therefore, in the design of our noise-rejection strategy, we focus on maximizing SKR, which also incorporates QBER. 
\subsection{Best-$K$ Strategy}
We first consider the optimal noise-rejection solution that fully exploits the knowledge of channel state information, i.e., transmittance and spatial probability values. Due to the non-uniform intensity distribution, detection is concentrated on elements with higher signal probability. In the Best-$K$ solution, the receiver selects the \(K\) elements with the largest value of \(P_i\) to achieve the maximum SKR, which is \(\hat{\mathbf  S}={\rm argmax}\{\mathrm{SKR}(\mathbf S)\}\). This strategy captures the region where the photon arrival probability is most concentrated, thus directly improving the signal contribution from the selected elements. Under uniform noise, selecting elements with higher spatial probability effectively increases the probability that a detected event is triggered by a signal rather than noise. But there is an optimal $K$ beyond which the remaining SPDs can be considered as noise dominated. It should be noted that, in the Best-\(K\) strategy, the value of \(K\) will be adaptively determined for each channel realization. 

%The Best-\(K\) solution serves as an upper-bound benchmark, as it requires a global ranking for the spatial probability for the whole array. 

\subsection{Global Threshold Strategy}
Our second proposed solution is based on determining a constant threshold value and has the potential to be implemented based on in-chip processing. To realize this solution, we consider a threshold-based selection strategy without requiring a detector ranking process. Instead of selecting a flexible subset for each channel realization, the receiver selects all elements with spatial probability exceeding a fixed threshold that does not vary over different coherence times. Therefore, the selected set is defined as \({\mathbf S(\tau)}=\{i: P_i\geq\tau\}\), where \(\tau\) is a constant global threshold that can be optimized based on the statistics of the channel before QKD operation. The optimal value of the threshold can be obtained as $\hat\tau={\rm argmax}\{\mathrm{SKR}(\mathbf S(\tau))\}$ and finally the optimal set is given by \(\hat{\mathbf S}=\{i: P_i\geq\hat\tau\}\). 

This scheme retains elements with relatively strong signal levels and effectively improves the performance of the system. Compared to the Best-\(K\) solution, the global threshold approach offers a potential reduction in computational complexity and easier application while still utilizing the spatial degree of freedom of detector arrays. However, the threshold is fixed through different channel realizations, thus preventing it from fully adapting to the spatial fluctuations, and making it suboptimal compared to the Best-$K$ strategy.  

\section{Numerical Result}
In this section, numerical simulations are presented to evaluate the performance of the proposed adaptive receiving strategies under different channel conditions. We consider a $8\times8$ detector array (i.e., $N=64$) and the wavelength of the light is assumed to be \(850\,\mathrm{nm}\).
 %In this section, we explain how we get the statistical characteristics of the spatial field intensity on the focal plane. Then, we compare the QBER and SKR performance of the Best-\(K\) strategies and global threshold strategies under different channel conditions.

\subsection{Statistical Characteristics of Spatial Intensity}
Using SSFM simulation of the wave propagation through the satellite-ground channel, we first obtain realizations of optical field at the receive aperture. Then, the field in the receiver’s focal plane, where the detector array is located, is obtained by taking the two-dimensional Fourier transform of the aperture field of receiver lens. The intensity of each detector element is obtained by integrating the intensity over its area in the focal plane. Through the Monte-Carlo method, we repeat the stimulation process under different realizations of the channel, demonstrating the statistical distribution of the intensity of each element.

Spatial intensity fluctuations in the focal plane are described by speckle patterns generated due to wavefront distortion induced by turbulence \cite{b19}. Theoretically, since a single SPAD element integrates through its area, the intensity of an element can be viewed as a superposition of multiple speckle modes if the speckle size is significantly smaller than the detector size. Thus, the intensity can be modeled as the sum of several random variables. As a result, when a sufficient number of speckle modes are aggregated, each speckle in the focal plane follows an exponential distribution\cite{b20}, therefore, the element intensity follows a Gamma distribution \cite{b20}. The parameter of the distribution depends on the turbulence conditions and optical configuration.  

\begin{figure}[!t]
\centering
\setlength{\tabcolsep}{2pt}

\begin{tabular}{cc}

\hspace{-2em}\includegraphics[width=0.58 \columnwidth ]{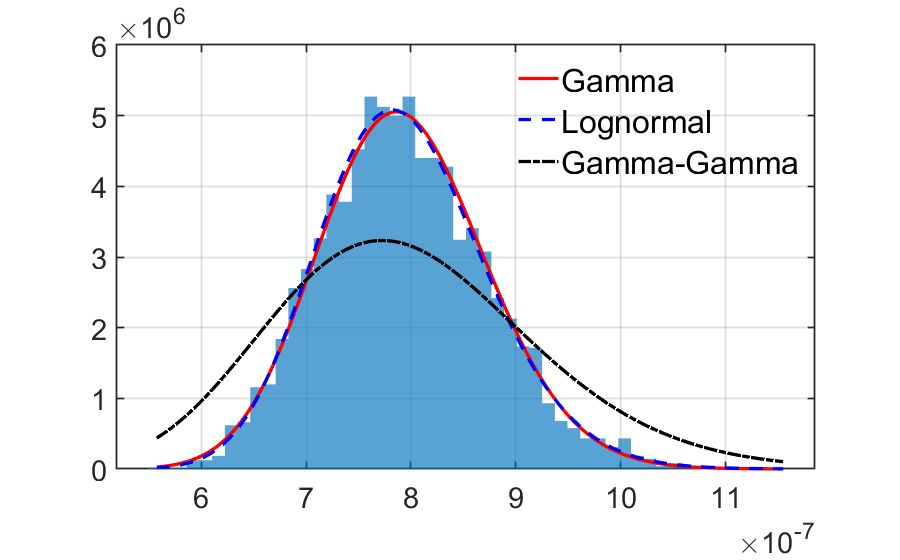} &\hspace{-2.5em}
\includegraphics[width=0.58\columnwidth]{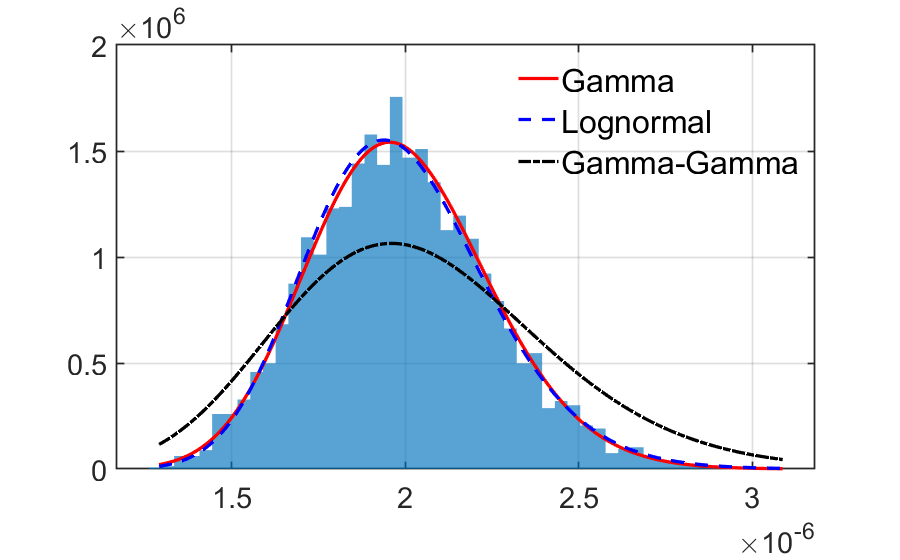} \\[-1em]
{\scriptsize (a)} & {\scriptsize (b)} \\[-0.2em]

\hspace{-1em}\includegraphics[width=0.6\columnwidth]{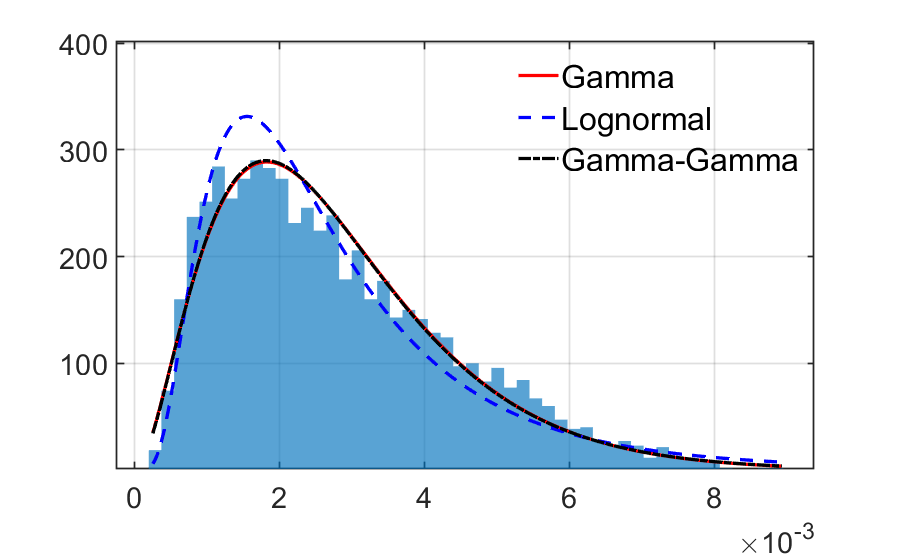} &\hspace{-2em}
\includegraphics[width=0.36\columnwidth]{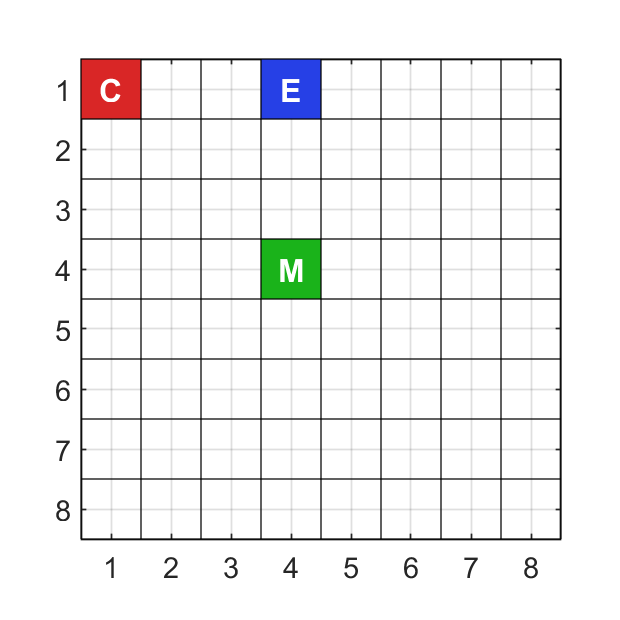} \\[-0.5em]
{\scriptsize (c)} & {\scriptsize (d)}

\end{tabular}

\caption{Statistical characterization of focal-plane intensity at different detector positions. 
(a) Corner element, (b) edge element, and (c) center element show the histogram fitting results. 
The horizontal axis represents the received intensity, and the vertical axis represents the occurrence frequency (histogram counts). 
(d) Schematic illustration of detector positions within the array.}
\label{fig:histogram_position}

\end{figure}

To characterize the statistical property of the spatial intensity distribution generated by our simulation, the intensity distributions are fitted using candidate models including the Gamma, lognormal, and Gamma–Gamma distributions. The goodness of fit is evaluated using the Kolmogorov–Smirnov (KS) test. The fitting results, as seen in Fig. \ref{fig:histogram_position} show that the optimal distribution depends on the spatial location and the underlying speckle structure. When a detector element integrates over many speckles, the intensity is well approximated by a Gamma distribution; with only a few dominant speckles, the distribution becomes more skewed and is better approximated by a lognormal model, consistent with the observed differences between center and corner elements under weak turbulence. Therefore, no single model is universally optimal, and Gamma and lognormal distributions exhibit complementary behavior across conditions. Note that our fit goodness tests match the observations above. These fitted models can be used for analytical performance analysis of the QKD systems, which is beyond the scope of this paper where the performance evaluation is conducted directly using simulated channel realizations.%, the Akaike Information Criterion (AIC) test, and the Bayesian Information Criterion (BIC) test

\begin{figure}[!t]
\centering
\setlength{\tabcolsep}{2pt}

\begin{tabular}{cc}

\includegraphics[width=0.5\columnwidth]{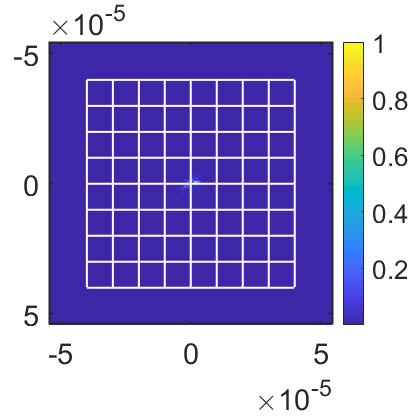} &
\includegraphics[width=0.5\columnwidth]{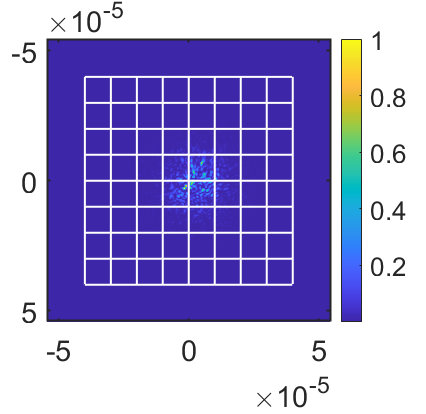} \\[-0.6em]
{\scriptsize (a)} & {\scriptsize (b)} \\[-0.2em]

\includegraphics[width=0.5\columnwidth]{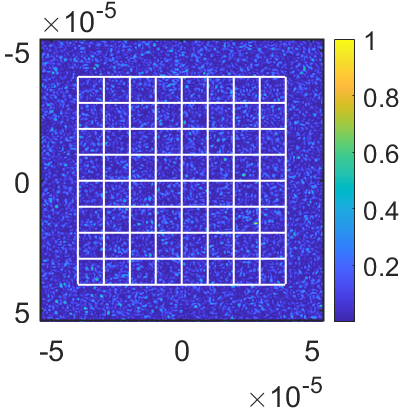} &
\includegraphics[width=0.52\columnwidth]{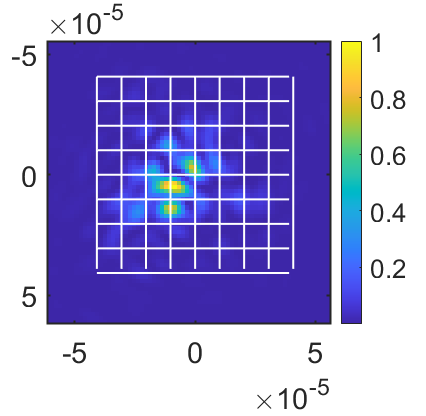} \\[-0.6em]
{\scriptsize (c)} & {\scriptsize (d)}

\end{tabular}

\caption{Example realizations of focal-plane speckle intensity distributions under different turbulence conditions. 
(a) Weak, (b) moderate, and (c) strong turbulence for a focal length of $f=0.5\,\mathrm{m}$, and (d) weak turbulence with an increased focal length of $f=4\,\mathrm{m}$. 
The horizontal and vertical axes represent spatial coordinates on the focal plane, and the color scale indicates intensity. 
The white grid denotes the detector array layout.}
\label{fig:speckle_realizations}

\end{figure}

As shown in Fig. \ref{fig:speckle_realizations}, the spatial intensity of the focal plane undergoes a significant transition with increasing turbulence strength. Under weak turbulence, the optical field remains relatively concentrated around the center and is dominated by the diffraction effect. With an increase in the turbulence strength, the spatial distribution becomes more spread out due to the phase distortion. In a strong turbulence scenario, the intensity becomes widely distributed, exhibiting a nearly random spatial pattern. This transition reflects the shift of the channel from being dominated by diffraction to turbulence. %The spatial degree of freedom increases during this process.
It should be noted that the above observations were obtained under a fixed focal length setup. The focal length plays an important role in the spatial intensity in the focal plane. Fig.~\ref{fig:speckle_realizations} indicates that different focal lengths will result in different characteristics of spatial intensity in the focal plane, which in turn influence the efficiency of the noise-rejection strategy. Further studies of this effect will be considered in future work. 

\subsection{Performance of Noise-Rejection Strategies}
The performance of the noise-rejection strategies is evaluated under different turbulence strengths and noise levels. We compared the two detection strategies proposed in this paper with two non-adaptive baseline schemes that activate either four central elements of the array or all elements of the array. The results show that the performance improvement strongly depends on the spatial characteristics of the receiver. 

\begin{figure}[!t]
\centering
\setlength{\tabcolsep}{2pt} 

\begin{tabular}{cc}

\hspace{-0.4em}\includegraphics[width=0.53\columnwidth]{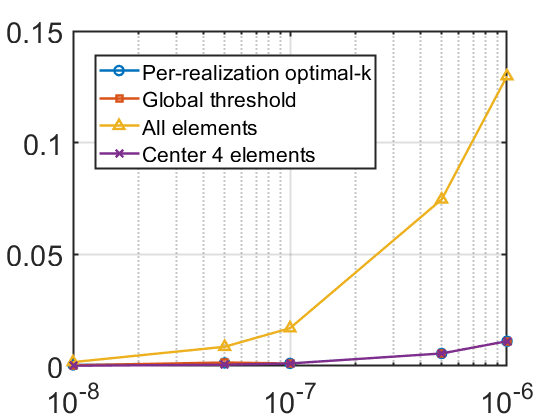} &\hspace{-1.2em}
\includegraphics[width=0.53\columnwidth]{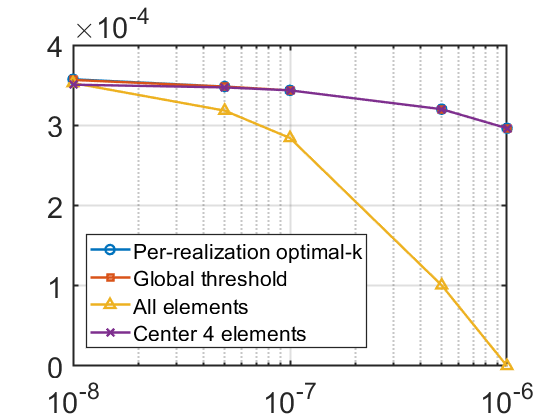} \\[-0.6em]
{\scriptsize (a)} &\hspace{-1em} {\scriptsize (b)} \\[-0.2em]

\hspace{-0.4em}\includegraphics[width=0.53\columnwidth]{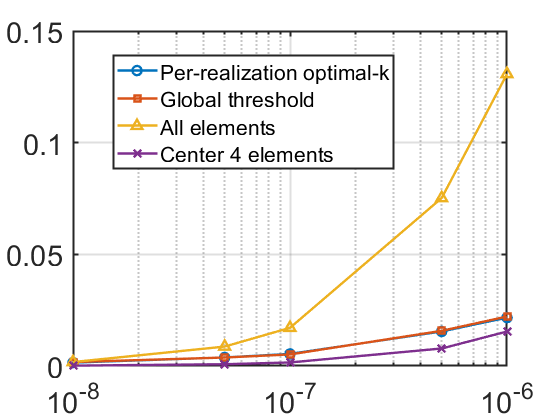} &\hspace{-1.2em}
\includegraphics[width=0.53\columnwidth]{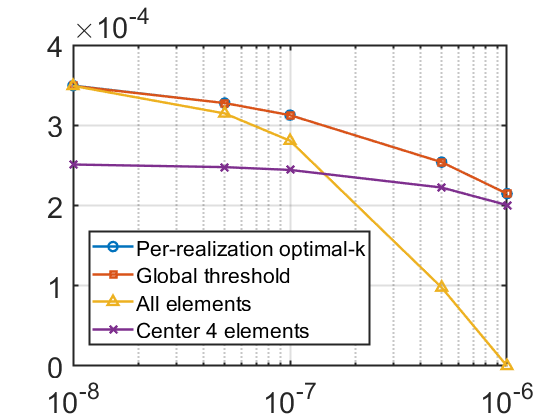} \\[-0.6em]
{\scriptsize (c)} &\hspace{-1em} {\scriptsize (d)} \\[-0.2em]

\hspace{-0.4em}\includegraphics[width=0.53\columnwidth]{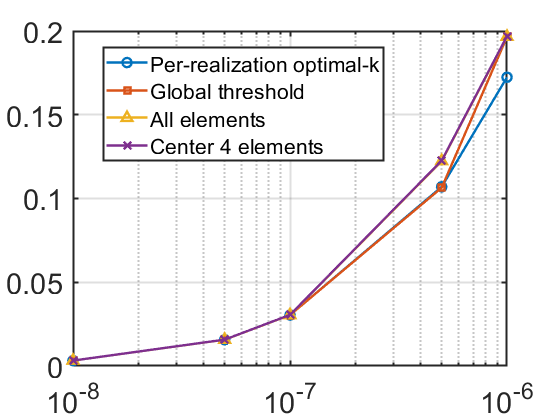} &\hspace{-1.2em}
\includegraphics[width=0.53\columnwidth]{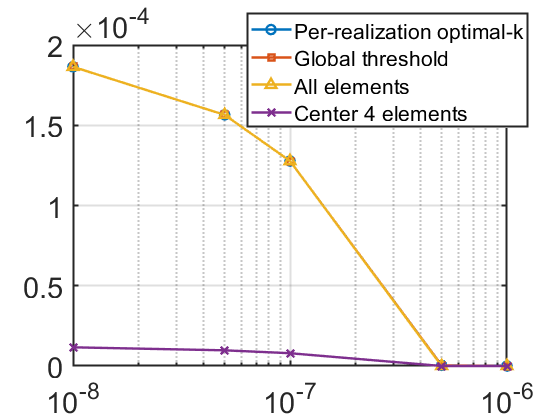} \\[-0.6em]
{\scriptsize (e)} & \hspace{-1em}{\scriptsize (f)}

\end{tabular}

\caption{Performance comparison of different detector-selection strategies under different turbulence conditions. 
(a) QBER and (b) SKR under weak turbulence; 
(c) QBER and (d) SKR under moderate turbulence; 
(e) QBER and (f) SKR under strong turbulence. 
The horizontal axis represents the noise level $n_B$, and the vertical axis denotes QBER or SKR (Hz), respectively.}
\vspace{-0.8em}
\label{fig:qber_skr_comparison}

\end{figure}
Under weak turbulence, the received optical field is highly concentrated at the center of the focal plane. The signal has a high probability of being concentrated over a small number of central elements. Therefore, the noise-rejection strategies have only a limited effect. On the other hand, under strong turbulence, the spatial intensity becomes highly dispersed and approaches a nearly uniform distribution on the focal plane. As a result, the differences between elements become smaller, which still reduces the impact of noise-rejection strategies. However, under moderate turbulence conditions, the performance gain of noise rejection is significant. In this regime, the spatial distribution between different detector elements is most pronounced. This phenomenon provides favorable conditions for the noise-rejection strategies, allowing the receiver to effectively choose the regions dominated by the signal while suppressing the noise effect.

The results indicate that the effectiveness of the noise-rejection strategies depends on the spatial distribution of the received signal. Only limited gains can be obtained under weak and strong turbulence, while significant improvement can be seen under moderate turbulence. In addition, as shown in Fig. \ref{fig:qber_skr_comparison} the global threshold strategy, as a sub-optimal solution, achieves close performance to the Best-\(K\) strategy, demonstrating its effectiveness in practice. Preliminary results also suggest that the performance varies by the focal length of the receiving optics, indicating that the focal length has the potential to serve as a tunable parameter in practical applications.  

\section{Conclusion}
In this paper, we investigate exploiting the spatial degrees of freedom SPD arrays for receiver noise rejection in SatQKD systems over atmospheric turbulence channels. The proposed noise-rejection strategy activates signal-dominant detector elements adaptively to enhance the performance of the system. The results show that the effectiveness of the adaptive noise-rejection strategy is highly dependent on the turbulence conditions and significant improvement is observed in moderate turbulence. For future work, we will further investigate the impact of focal length on the effectiveness of the proposed strategies. In addition, mode refined modeling of the spatial intensity distribution will be explored to improve the accuracy of the statistical characteristic of the receiver plane. 

\section*{Acknowledgment}
This work was supported by the UK Engineering and Physical Sciences Research Council (EPSRC) through QUANTA: Quantum Communications over Atmospheric Channels Aided by Detector Arrays under grant UKRI2875.

\bibliographystyle{ieeetr}
\bibliography{IEEEexample}

\end{document}